\documentclass{amsart} 
\usepackage[margin=1in]{geometry}
\usepackage{float}
\usepackage{color}
\usepackage{graphicx}
\usepackage[labelformat=simple]{subcaption}

\usepackage{mwe}
\usepackage{listings}
\definecolor{dkgreen}{rgb}{0,0.6,0}
\definecolor{gray}{rgb}{0.5,0.5,0.5}
\definecolor{mauve}{rgb}{0.58,0,0.82}
\usepackage{gensymb}
\usepackage{hyperref} 
\begin{document}
\title{A Mathematical Model For Meat Cooking}
\author{H. AH Shehadeh$^{3*}$}
\author{S. Deyo$^1$}
\author{S. Granzier-Nakajima$^{2}$}
\author{P. Puente$^2$}
\author{K. Tully$^4$}
\author{J. Webb$^3$}
\thanks{* Corresponding author. E-mail address alhajjhy@jmu.edu \\
$^1$University of Florida, Gainesville, FL, USA \\
$^2$University of Arizona, Tucson, AZ, USA \\
$^3$James Madison University, Harrisonburg, VA, USA\\
$^4$ Wheaton College, Wheaton, IL, USA}
\begin{abstract}
We present an accurate two-dimensional mathematical model for steak cooking based on Flory-Rehner theory. The model treats meat as a poroelastic medium saturated with fluid. Heat from cooking induces protein matrix deformation and moisture loss, leading to shrinkage. Numerical simulations indicate good agreement with experimental data. Moreover, this work presents a new and computationally non-expensive method to account for shrinkage. \\
\smallskip \\
\noindent \textbf{Keywords: Flory-Rehner Theory, Meat Cooking, Poroelastic Medium, Darcy's Law}
\end{abstract}
\maketitle
\vspace{-2em}\section{Introduction}\label{sec1}
Mathematical modeling of the cooking process of meat is vital to understand the underlying physical phenomena and to help improve safety of consumed meat, as well as optimize the quality and flavor of meat and extend its shelf-life (see \cite{FoodProcess, R4} and the references therein for highlighting the importance of mathematically modeling the thermal cooking of meat and accurately describing the involved physical processes). 

In this paper, we model meat as a fluid-saturated poroelastic medium composed of a solid matrix (polymer) and fluid. While traditional poroelastic theory considered solid phase non-hygroscopic, modern approaches include biphasic and monophasic approaches (see for example  \cite{R5} modeling polymeric gels, \cite{R55} modeling spherical hydrogels and \cite{R555} modeling cartilage).

As the temperature increases during cooking, a pressure gradient builds and induces fluid motion and deformation of the solid matrix, which also contributes to the motion of the fluid. The model accounts for temperature distribution, fluid velocity field, moisture content, surface evaporation, and shrinkage during the cooking process.

Particular types of meat shrink at different rates. Leaner cuts generally shrink less because they contain less water and fat. According to the U.S. Department of Agriculture, beef sirloin and brisket shrink by around $16\%$ and $30\%$, respectively. A broiled chicken wing may lose only around $14\%$ of its weight, while a whole roast chicken may shrink by nearly $40\%$. Cured roast ham may shrink by $8\%$, while a grilled pork patty could lose $30\%$ of its weight \cite{everydaylife}. Experiments indicate that during double-sided pan frying of beef burgers, the pressure-driven water loss (up to $80\%$ of the water loss) is a much more important mechanism governing the water loss than the evaporation losses occurring at the surface. Fat losses, in the form of drip, increase significantly with fat content and are not significantly influenced by the cooking temperature \cite{tornberg2013engineering}.

Browsing the literature attempting to mathematically model meat cooking, we find \cite{R4} which considers a polymer solvent model for steak cooking which includes additional friction forces between the polymer and the solvent in their force balance equations. However, these forces act in equal and opposite directions on the polymer and the solvent, so they cancel out and have no effect. The authors also ignore viscous forces, elastic effects, and shrinkage. The simulations in \cite{R4} are performed in one dimension using finite differences and simplified boundary conditions. We do however follow \cite{R4} in deriving the governing equations from fundamental principles: conservation of mass of the fluid, balance of forces on the fluid, and conservation of energy.

To account for shrinkage during cooking, it is often considered that the change of dimensions is proportional to the volume of fluid loss \cite{du2005correlating, katekawa,SUN-HU}. While heat and mass transfer in solid food was incorporated in \cite{R4, van2007soft}, they neglect shrinkage. In \cite{fowler1991effect}, the authors consider shrinkage as the integrated result of temperature dependent and volumetrically-distributed shrinking. Accounting for meat shrinkage decreases the predicted cooking time compared to the classical Heisler chart for conduction in a constant-volume cut. However, the model only considers heat transfer and simulations are performed in one dimension. In \cite{zorrilla2003heat}, the authors investigate heat transfer in double-sided cooking of meat patties in two dimensions with radial shrinkage using finite differences and compare it to a one-dimensional model. While the temperature predictions for the geometric center of a patty are similar, the two-dimensional model monitors temperature variation at regions near the circumferential edge of the patty. A finite element approach is presented in \cite{goni2010prediction} with irregular geometries as simulation domains. Heat transfer is modeled by Fourier's law, while the change in internal moisture content is modeled as a function of water demand. Predictions of evaporative loss, dripping loss, and cooking time match well with experimental results.
\section{Model Assumptions}
We model steak as a poroelastic polymer medium (protein network or matrix) saturated with water, though in reality, the pores are filled with an aqueous solution of plasma and ions and other soluble proteins. Fluid in the steak includes ``bound water" held in the protein matrix, and ``free water" held in capillaries in between muscle fibers that can escape from the boundary of the steak into the environment \cite{Bertram, VS-RGM}. Throughout this paper, we will use two distinct but related notions to discuss quantities of fluid present in the steak medium: porosity and moisture content, which we will now define.

Porosity is defined as a volume fraction:
\begin{equation}
1-\phi(x_0,y_0;t_0)=\lim_{r\to 0} \frac{V_{f, (x-x_0)^2+(y-y_0)^2\le r^2}}{\pi r^2},
\label{eqn:porosity}
\end{equation}
where $V_{f, (x-x_0)^2+(y-y_0)^2\le r^2}$ is the volume of free fluid inside a disk of radius $r$ and $\phi(x_0,y_0;t_0)$ is understood as the volume ratio of the solid protein matrix.


In comparison, moisture content is defined as a mass ratio:
\begin{equation}
n=\frac{(1-\phi)\rho_f}{(1-\phi)\rho_f+\phi \rho_s},
\label{eqn:Moisture}
\end{equation}
where $\rho_s$ and $\rho_f$ are the mass densities of the solid protein matrix and fluid, respectively.

We assume the following:
\begin{enumerate}
\item The steak is lean ($\leq 4\%$ fat).
\item Pores are fully saturated with fluid.
\item The density, $\rho_f$, is constant.
\item Protein matrix is ordered with little crosslinking.
\end{enumerate}
\section{Governing Laws and Equations of Motion}
The governing equations are derived from fundamental principles: conservation of mass of the fluid, balance of forces on the fluid, and conservation of energy.  
\subsection{Conservation of mass}
Let $A$ be a small area in the interior of the domain $\Omega$, with boundary $\partial A$, and let ${\bf n}$ be the outward unit normal vector. The mass flux of the fluid is defined by ${\bf J}=\rho_f (1-\phi) {\bf w}$, where $\rho_f$ is the density of the fluid, $1-\phi$ is the fluid volume fraction, and ${\bf w}$ is the local fluid velocity.

The rate of change of mass $m$ inside the domain $A$ accounts for the gain and loss of mass ($m_{in}-m_{out}$) through the boundary $\partial A$, thus 
\begin{align*}
\frac{\partial m}{\partial t} & = -\oint_{\partial A} \mathbf{J} \cdot \mathbf{n} ds\\
				& = -\int_A \nabla \cdot \mathbf{J} dS\\
				& = -\int_A \nabla \cdot \rho_f (1-\phi) {\bf w} dS.
\end{align*}
Using the relationship between mass and density, we have
\begin{align*}
\frac{\partial}{\partial t}\int_A (\rho_f (1-\phi)) dS = -\int_A \nabla \cdot \rho_f (1-\phi) {\bf w},
\\
\frac{1}{area(A)}\int_A \frac{\partial}{\partial t} (\rho_f(1-\phi)) + \nabla \cdot \rho_f (1-\phi) {\bf w} = 0.
\end{align*}
The above expression is valid for all $A$ away from the boundary. Taking the limit as $area(A) \to 0$, we get the mass conservation equation for the fluid:
\begin{equation}
\frac{\partial}{\partial t} (\rho_f(1-\phi)) + \nabla \cdot \lbrace\rho_f (1-\phi) \mathbf{w}\rbrace=0.
\end{equation}
Assuming $\rho_f$ is constant, we find
\begin{equation}
\phi_t = \nabla \cdot ((1-\phi) \mathbf{w}).+ D_{w}\nabla^2 \phi.
\label{eqn:continuity}
\end{equation}

There is debate over whether moisture transport is a diffusive phenomena \cite{van2007soft}, and thus we simulate our model with and without water diffusion. To simulate with diffusion, we generously estimate $D_w \approx 7\times 10^{-9}m^2/s$ at $100^\circ C$ using room temperature experimental values of the diffusion coefficient for meat and the Arrhenius equation, which states that $D_w$ scales exponentially in temperature \cite{diffusion}. However, in the rest of this paper we set $D_w =0$, and only report on $D_w\neq 0$ in the conclusion.
\subsection{Balance of forces} 
This is usually referred to as conservation of momentum. Motivated by \cite{R4}, we model moisture transport in cooking meat with the Flory-Rehner theory of swelling or shrinking polymer gels. Note that Flory-Rehner theory assumes isotropic and uniform deformation which is not the case in our model, so we make this approximation while acknowledging this shortcoming of the model. 

Disregarding electrostatic interactions, the swelling pressure, $\pi_{sw}$, is expressed as $\pi_{sw}=\pi_{mix}+\pi_{el}$, where $\pi_{mix}$ is the osmotic pressure due to the change in entropy of mixing of solid matrix protein polymers with fluid monomers and $\pi_{el}$ is the network pressure which opposes the osmotic pressure. From Flory-Rehner theory, $\pi_{mix}$ and $\pi_{el}$ are derived via free energy expressions in the following form:
\begin{equation}
	\pi_{mix} = \frac{\partial F_{mix}}{\partial \phi} = \frac{RT}{V_f} \bigg[\ln\left(1-\phi\right)+\phi + \chi(T,\phi)\phi^2\bigg],
\label{eqn:pi_mix}
\end{equation}
\begin{equation}
	\pi_{el} =\frac{\partial F_{el}}{\partial \phi} = \frac{RT\rho_s}{M_c}\bigg[\phi^{1/3}\phi_0^{2/3}-\frac{1}{2}\phi \bigg],
\label{eqn:pi_el}
\end{equation}
where $\phi$ is the polymer volume fraction, $T$ is the absolute temperature, $R$ the gas constant, $\chi(T,\phi)$ is the Flory-Huggins interaction parameter, $V_f$ is the molar volume of the solvent, $\rho_s$ is the density of the polymer, $M_c$ is the average molar weight of polymer segments in between crosslinks, and $\phi_0$ the polymer volume fraction at crosslinking. 

In accordance with universal scaling laws investigated in \cite{van2013modeling}, the Flory-Huggins interaction parameter is
\begin{equation}
\chi(T, \phi) = \chi_p(T)-\left(\chi_p(T)-\chi_0\right)(1-\phi^2),
\label{eqn:FH_interation}
\end{equation}
where $\chi_p(T)$ is empirically fit using the sigmoidal function
\begin{equation}
\chi_p(T)=\chi_{pn}-\frac{\chi_{pd}-\chi_{pn}}{1+A\exp(\gamma[T-T_e])}.
\label{eqn:FH_sigmoidal}
\end{equation}
\begin{figure*}[!h]
\includegraphics[width=.5\textwidth]{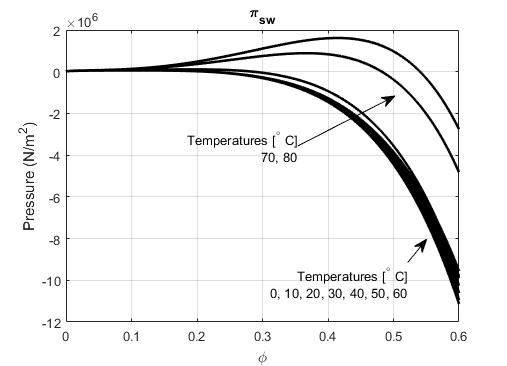}
\caption[]%
            {{$\pi_{sw}$ vs. $\phi$ for Temperatures 0 - 80 $^\circ C$}} 
\label{fig:pi_sw}
\end{figure*}
\\
Using Darcy's law, the momentum balance is
$$\rho_f (1-\phi){\bf w}=-\frac{\kappa}{\eta}\nabla \pi_{sw}.$$
Equivalently, via the relationship $\eta=\mu/\rho_f$ between kinematic viscosity $\eta$ and dynamic viscosity $\mu$,
\begin{equation}
(1-\phi){\bf w}=-\frac{\kappa}{\mu}\nabla \pi_{sw},
\label{eqn:momentum}
\end{equation}  
where $\rho_f$ is the bulk density of water, $\mathbf{w}$ is the fluid velocity, $\kappa$ is the permeability tensor, and $\mu$ is the dynamic viscosity. As in \cite{viscosity}, we model the temperature dependence of viscosity with the following formulation:  
\[\mu(T) = 2.414\times10^{-5} \times 10^{247.8/(T - 140)}.\]

We orient the steak such that the grain is in the $y$-direction. Since the permeability parallel to the grain empirically has been found to be $1.16$ times greater than it is perpendicular to the grain (i.e., $x$-direction) \cite{SUN-HU}, our permeability tensor is
\begin{equation*} 
 \kappa =    \begin{bmatrix} 
      \kappa^{(11)} & \kappa^{(12)} \\
      \kappa^{(21)} & \kappa^{(22)} 
   \end{bmatrix}
 = \begin{bmatrix} 
      \frac{1}{1.16} & 0 \\
      0 & 1
   \end{bmatrix}\kappa_{||}, 
\end{equation*}
where $\kappa_{\parallel}$ is the permeability parallel to the fibers. Permeability of food solids is not well known. While \cite{datta2006hydraulic} estimates that the permeability of food tissues is between $10^{-17}$ and $10^{-19}$ m$^2$, ground beef has been found to have permeability as high as $10^{-15}$ $m^2$ \cite{Feyissa}. Following \cite{Feyissa}, we consider a range of possible permeability coefficient values between $10^{-16}$ and $10^{-17}$ $m^2$. In the simulations, we also consider variable permeability fitted sigmoidally between $2.5\times 10^{-18}$ and $2.5\times 10^{-17} m^2$.
\subsection{Conservation of energy}
Heat is transferred through the poroelastic medium by conduction and convection (with the velocity of the fluid). Hence,
\begin{equation}
(cT)_t+\nabla \cdot (\rho_fc_f{\bf w}(1-\phi)T)=\nabla \cdot (k \nabla T),
\label{eqn:AdvDiff}
\end{equation} 
where $c$ is the effective specific heat capacity and $k$ is the effective anisotropic thermal conductivity:
\begin{equation}
\begin{aligned}
c&=\phi \rho_sc_s+(1-\phi)\rho_f c_f,\\
k&=\begin{bmatrix}
	k_{11} & 0 \\
    0 & k_{22} 
\end{bmatrix},\\
k_{22} &= \phi k_s+(1-\phi)k_f,\\
\frac{1}{k_{11}} &=\frac{1-\phi}{k_f}+\frac{\phi}{k_s},\\
\end{aligned}
\label{eqn:CandK}
\end{equation}
where the subscripts $s$ and $f$ denote the solid and fluid respectively, and $k_{11}$ and $k_{22}$ denote the heat conductance perpendicular and parallel to the fibers. We use the series and parallel resistor models to obtain the perpendicular and parallel heat conductance. Note that $c$ and $k$ are functions of temperature $T$ because $\phi$ depends on $T$. 
\subsection{Complete model}
For $(x,y;t) \in \Omega(t)$, the unknown quantities are:
\begin{enumerate}
\item Porosity: $1-\phi(x,y;t)$,
\item Fluid velocity: ${\bf w}(x,y;t)$, and
\item Temperature: $T(x,y;t)$.
\end{enumerate}
The governing equations are:
\begin{enumerate}
\item \textbf{Mass balance:} $\phi_t = \nabla \cdot ((1-\phi) \mathbf{w})$
\item \textbf{Conservation of momentum:}
$(1-\phi)\mathbf{w}=\kappa\nabla\cdot \left((1-\phi)\frac{\nabla {\bf w}+\nabla \mathbf{w}^T}{2}\right)-\frac{\kappa}{\mu}\nabla \pi_{sw}$, where
$$\begin{aligned}
\pi_{sw}&=\pi_{mix}+\pi_{el},\\
\pi_{el} &= \frac{RT\rho_s}{M_c}\bigg[ \phi^{1/3}\phi_0^{2/3}-\frac{1}{2}\phi\bigg],\\
\pi_{mix}&= \frac{RT}{V_f} \left[\ln(1-\phi)+\left(1-\frac{1}{n}\right)\phi + \chi(T,\phi)\phi^2\right],\\
\chi(T, \phi) &= \chi_p(T)-(\chi_p(T)-\chi_0)(1-\phi)^2,\\
\chi_p(T)&=\chi_{pn}+\frac{\chi_{pd}-\chi_{pn}}{1+A\exp(\gamma(T-T_e))}.
\end{aligned}$$
\item \textbf{Conservation of energy:}
$(cT)_t+\nabla \cdot (\rho_fc_f{\bf w}(1-\phi)T)=\nabla \cdot (k \nabla T)$, where
\begin{equation*}
c =\phi \rho_sc_s+(1-\phi)\rho_f c_f,
\end{equation*}
\begin{equation*}
k =\begin{bmatrix}
	k_{11} & 0 \\
    0 & k_{22} 
\end{bmatrix}
=\begin{bmatrix}
	\frac{k_s k_f}{(1-\phi)k_s+\phi k_f} & 0 \\
    0 & \phi k_s+(1-\phi)k_f 
\end{bmatrix}.
\end{equation*}
\end{enumerate}
\section{Boundary Conditions}\label{Sec:BCs}


\subsection{Porosity boundary conditions}
We use
\begin{equation}
\pi_{sw}=0
\label{eqn:phi_dirichlet}
\end{equation}
to determine $\phi$ on the boundary; that is, we choose the $\phi$ that ensures $\pi_{mix}+\pi_{el}=0$ on the surface of the steak. This condition implies that the surface is not pressurized.

\subsection{Fluid velocity boundary conditions}
When applying \eqref{eqn:momentum} to determine $\mathbf{w}$ on the boundary, we set $\pi_{sw}=0$ on the boundary. 

\subsection{Temperature boundary conditions}
Let $h_c$ denote the heat transfer coefficient, $r$ the latent heat of evaporation, and $j_{evap}$ the mass flux due to evaporation. We impose the condition
\begin{equation}
-k\frac{\partial T}{\partial n} = h_c(T-T_D)+rj_{evap}.
\end{equation}
The mass flux due to evaporation is given by
\begin{equation*}
j_{evap} = h\left(\sqrt[3]{\left(\frac{D_a}{k_a}\right)^2 \frac{1}{\rho_a c_a}}\right)\left[\frac{\left(1-\frac{X_m}{X}\right)p_{sat,0}\exp\left(17.27\frac{T-T_0}{T-35.86}\right)M_f}{RT}-c_0\right],
\end{equation*}
where $D_a$ denotes the diffusion coefficient of water in air, $k_a$ the thermal conductivity of air, $\rho_a$ the density of air, $c_a$ the specific heat of air, $M_f$ the molar mass of the fluid, and $c_0$ the water vapor concentration in the boundary layer \cite{van2007soft}. In empirical investigation, Bengston (1976) found that the wet-bulb temperature changes in the oven during cooking; and thus, in order to validate the model, $c_0$ is modeled to match the empirical wet bulb temperature \cite{Bengston}. The term $\left(1-\frac{X_m}{X}\right)$ approximates the water activity $a_w$ from the GAB model. The moisture content on a dry weight basis $X$ is given by 
\begin{equation*}
X=\frac{m_f}{m_s}=\left(\frac{1}{\phi}-1\right)\frac{\rho_f}{\rho_s}.
\end{equation*}
        \begin{table}[H] 
		\centering
        
		\caption{Model Variables and Parameters}
		\resizebox{.98\textwidth}{!}{\begin{tabular}{l l c c c}
			\hline\hline
            Variable & Description & Value & Unit & Source\\
            \hline
            $\phi(\mathbf{x};t)$ & Protein matrix volume fraction & --- & --- & --- \\
            $1-\phi(\mathbf{x};t)$ & Fluid volume fraction & --- & --- & --- \\
            $\mathbf{w}(\mathbf{x};t)$ & Fluid velocity & --- & m/s & --- \\
            $T$ & Temperature & --- & K & --- \\
            $T_0$ & Initial Temperature & --- & K & --- \\
            $T_D$ & Maximum Temperature on $\partial \Omega$ & --- & K & --- \\
            $\pi_{sw}(\mathbf{x};t)$ & Swelling pressure & --- & N/$m^2$ & --- \\
            $\chi(T,\phi)$ & Flory-Huggins temperature- and moisture-dependent interaction & --- & --- & --- \\
            $\chi_0$ & Flory-Huggins interaction for fully hydrated polymer & $0.5$ & --- & \cite{R4} \\
            $\phi_0$ & Polymer fraction at crosslinking & $.217$ & --- & \cite{van2007soft,initial_porosity} \\
            $\phi_{init}$ & Initial protein volume fraction & --- & --- & \\
            $\chi_p(T)$ & Flory-Huggins temperature-dependent interaction & --- & --- & --- \\
            $\chi_{pn}$ & Flory-Huggins interaction of dry solid protein matrix & $0.7$ & --- & --- \\
            $\chi_{pd}$ & Flory-Huggins interaction of denatured protein matrix& $0.9$ & --- & --- \\
            $X_m$ & Moisture content of the first monolayer of absorbed water & $0.08$ & --- & \cite{van2007soft} \\
            $X$ & Moisture content on dry weight basis & --- & --- & --- \\
            $A$ & $\chi_p$ sigmoidal least squares fitting parameter& $30$ & --- & \cite{R4,VanderSman2007} \\
            $\gamma$ & $\chi_p$ sigmoidal least squares fitting parameter& $-0.25$ & --- & \cite{R4,VanderSman2007} \\
            $T_e$ & $\chi_p$ sigmoidal least squares fitting parameter & $325$ & K & \cite{R4,VanderSman2007}\\
            $R$ & Gas constant & $8.314$ & J/molK & --- \\
            $p_{sat,0}$ & Empirical Tetens coefficient & $597$ & N/$m^2$ & \cite{van2007soft} \\
            $V_f$ & Molar volume of solvent & $1.8 \times 10^{-5}$ & $m^3$/mol & --- \\
            $M_c$ & Average molar weight of polymer segments between cross-links & $\approx$ 6 (Elastin) & kg/mol & \cite{Ertbjerg,van2007soft} \\
            $M_f$ & Molar weight of fluid & $18 \times 10^{-3}$ & kg/mol & --- \\
            $\eta$ & Kinematic viscosity of fluid & $10^{-6}$ & $m^2$/s & \cite{van2007soft} \\
            $\mu(T)$ & Dynamic viscosity of fluid & --- & Ns/$m^2$ & --- \\
            $c(\mathbf{x};t)$ & Effective specific heat & --- & J/$m^2$K & --- \\
            $c_{s}$ & Specific heat of solid protein matrix & $2008$ & J/kgK & \cite{R4,choi1986thermal} \\
            $c_{f}$ & Specific heat of fluid (solvent) & $4178$ & J/kgK & \cite{R4,choi1986thermal} \\
            $c_{a}$ & Specific heat of air & $1006$ & J/kgK &  \\
            $k(\mathbf{x};t)$ & Effective thermal conductivity & --- & W/mK & --- \\
            $k_s$ & Thermal conductivity of solid matrix & $0.18$ & W/mK & \cite{R4,choi1986thermal} \\
            $k_f$ & Thermal conductivity of fluid & $0.57$ & W/mK & \cite{R4,choi1986thermal} \\
            $k_a$ & Thermal conductivity of (dry) air ($20^{\circ}$C) & $0.02587$ & W/mK & \cite{choi1986thermal} \\
            $k_{22}$ & Thermal conductivity parallel to fibers & $0.500$ & W/mK & \cite{van2007soft,initial_porosity} \\
            $k_{11}$ & Thermal conductivity perpendicular to fibers & $0.404$ & W/mK & \cite{van2007soft,initial_porosity} \\
            $\kappa_{\parallel}$ & Permeability parallel to fibers & $\approx 10^{-17},10^{-16}$ or $2.5\times 10^{-18}\text{ to }2.5\times10^{-17}$ & $m^2$ & \cite{datta2006hydraulic} \\
            $\kappa_{\perp}$ & Permeability perpendicular to fibers & $\frac{1}{1.16}\kappa_{\parallel}$ & $m^2$ & \cite{van2007soft,SUN-HU} \\
            $\rho_s$ & Density of protein matrix & $1300$ & kg/$m^3$ & \cite{van2007soft,gosline1978temperature} \\
            $\rho_f$ & Density of fluid & $1000$ & kg/$m^3$ & \cite{R4,choi1986thermal} \\
            $\rho_a$ & Density of air & $1.225$ & kg/$m^3$ & --- \\
            $h_c$ & Heat transfer coefficient & $\approx 15$ & W/$m^2$K & \cite{Willix} \\
            $r$ & Latent heat of evaporation & $2.257 \times 10^6$ & J/kg & --- \\
            $j_{evap}(\mathbf{x};t)$ & Mass flux due to evaporation & --- & kg/$m^2$s & --- \\
            $c_0$ & Vapor concentration in boundary layer ($20^\circ$C; $50\%$ humidity) & $0.0087$ & kg/$m^3$ & --- \\
            $c_0$ & Vapor concentration in boundary layer ($50^\circ$C; $100\%$ humidity) & $0.083$ & kg/$m^3$ & --- \\
            $D_a$ & Diffusion coefficient of water vapor in air & $2.5 \times 10^{-5}$ & $m^2$/s & --- \\
            $D_w$ & Diffusion coefficient of water in meat & $7 \times 10^{-9}$ & $m^2$/s & \cite{diffusion} \\
            $M_f$ & Molar mass of fluid & $0.018$ & kg/mol & --- \\
                        $D$ & Thermal diffusivity of fluid & $1.4 \times 10^{-7}$ & $m^2$/s & --- \\
                                    $\nu = \frac{\rho_s c_s}{\rho_f c_f}$ & Non-dimensional parameter& $0.6248$ & --- & --- \\
            $\omega=\frac{k_s}{k_f}$ & Non-dimensional parameter& $0.3158$ & --- & --- \\
                        $\alpha = \frac{\nu T_0}{T_D-T_0}$ & Non-dimensional parameter& --- & --- & ---\\
            $\lambda =\frac{l^2r\rho_f}{k_s t_0(T_D-T_0)}$ & Non-dimensional evaporation coefficient& --- & --- & --- \\
            $\hat{c}=1-\phi(1-\nu)$ & Non-dimensional heat capacity & --- & --- & ---\\
            $\hat{k}_{22}=1-\phi(1-\omega)$ & Non-dimensional conductivity perpendicular to grain & --- & ---& ---\\
            $\hat{k}_{11}=\frac{\omega}{\omega(1-\phi)+1}$ & Non-dimensional conductivity parallel to grain & --- & --- & ---\\
            $\beta_{el}=\frac{t_0 R \rho_s T_0}{\mu_0 M_c}$ & Non-dimensional elastic pressure coefficient &--- & --- & ---\\
            $\beta_{mix}=\frac{t_0 R T_0}{\mu_0 V_f}$ & Non-dimensional osmotic pressure coefficient & --- & --- & ---\\
            $t_0=\frac{l^2}{D}$ & Time scale & $960$ & s & ---\\
            $l=\frac{k_s}{h_c}$ & Length scale& $0.012$  & m & ---\\
            $w_0 = \frac{l}{t_0}$ & Velocity scale & $1.25 \times 10^{-5}$ & m/s & ---\\
            $\pi_0=\frac{\mu_0}{t_0}$ & Pressure scale & $1.04 \times 10^{-6}$ & N/$m^2$ & ---\\
            $j_{evap_0} = \frac{\rho_f l}{t_0}$ & Evaporation scale & $0.0125$ & kg/$m^2$s & --- \\
\end{tabular}}
            \end{table}
            \pagebreak
\subsection{Moving boundary and shrinkage}
We divide our domain into discrete cells of size $h$ for the purpose of numerical simulation. We view shrinkage as local. Our coordinate system is fixed to the protein matrix, so the solid volume within each discrete cell is constant; i.e., all volume change is because of fluid entering or leaving. At constant temperature, we assume that each discrete cell maintains its aspect ratio regardless of fluid content. Hence,
\begin{equation}
h^2=v=v_s+v_f=\phi_{init}h_0^2+(1-\phi)h^2,
\end{equation}
where $\phi_{init}$ is the initial protein volume fraction, $h_0$ is the intitial step size, and $v$ denotes the volume of a discrete cell. Solving for $h$, we obtain
\begin{equation}
h=h_0\sqrt{\frac{\phi_{init}}{\phi}}
\end{equation}
as our step size for each discrete cell.

\begin{figure}
\minipage{0.32\textwidth}
  \includegraphics[width=\linewidth]{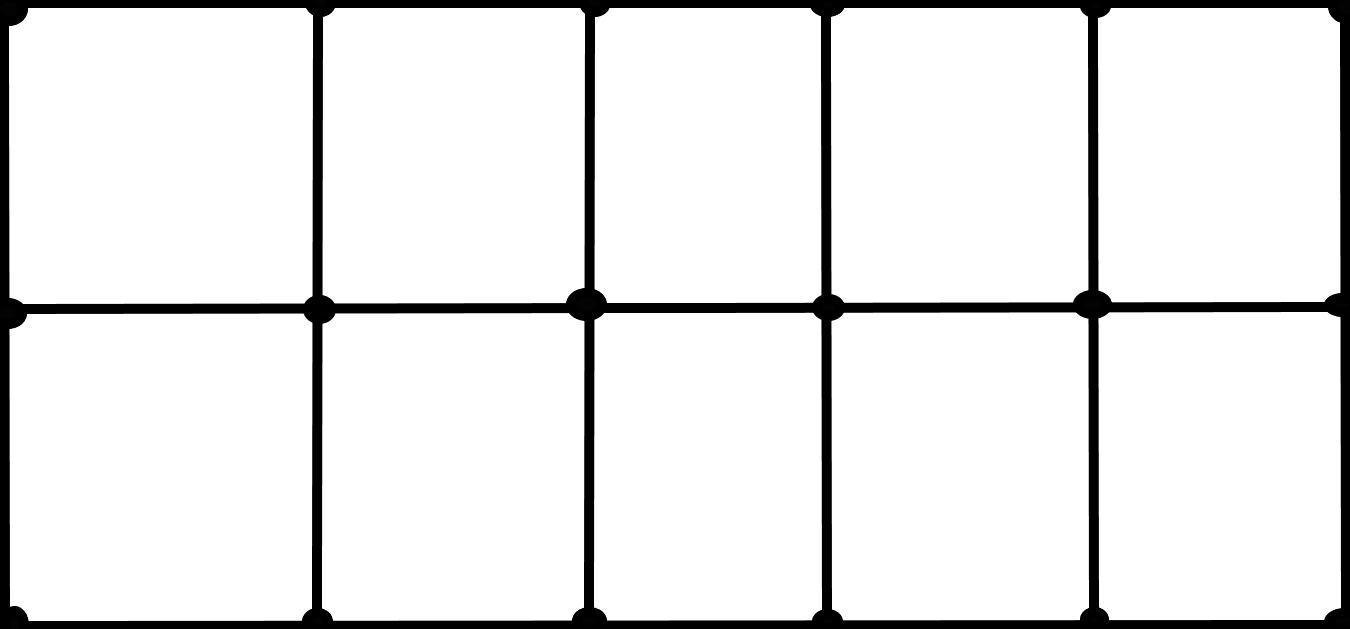}
  \caption{Pre-shrinkage.}\label{fig:pre-shrinkage}
\endminipage\hfill
\minipage{0.32\textwidth}
  \includegraphics[width=\linewidth]{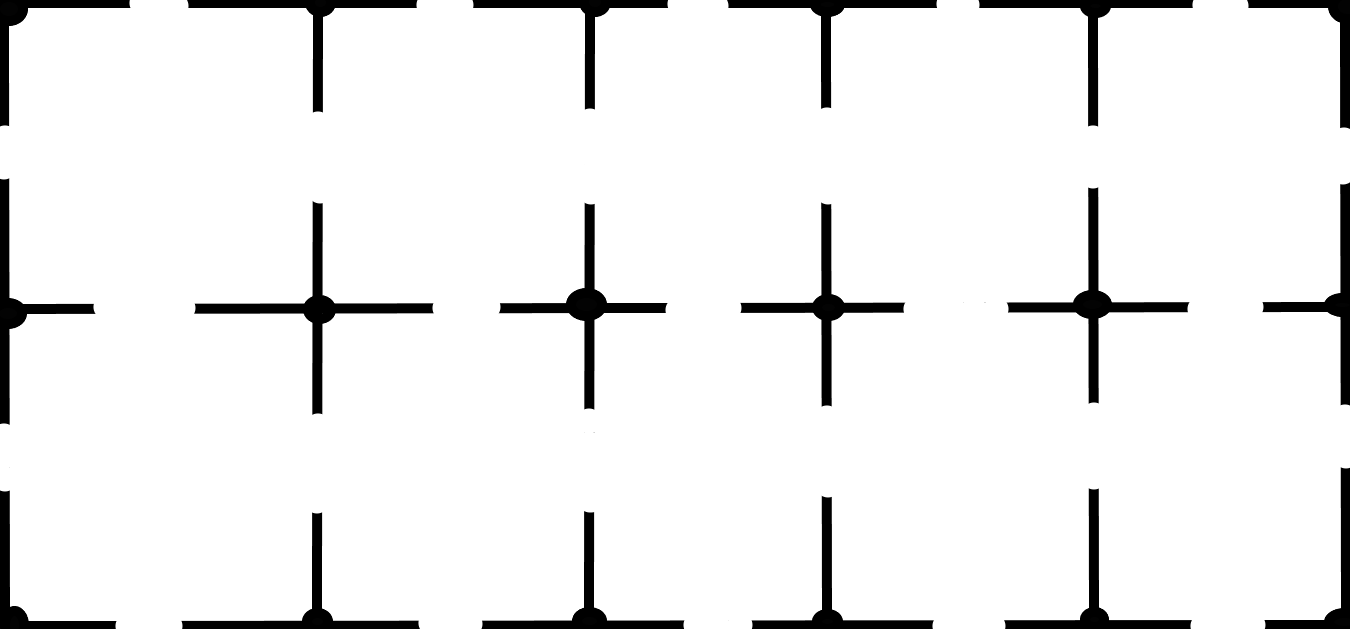}
  \caption{Post-shrinkage approximation.}\label{fig:post-shrinkage approximation}
\endminipage\hfill
\minipage{0.32\textwidth}%
  \includegraphics[width=\linewidth]{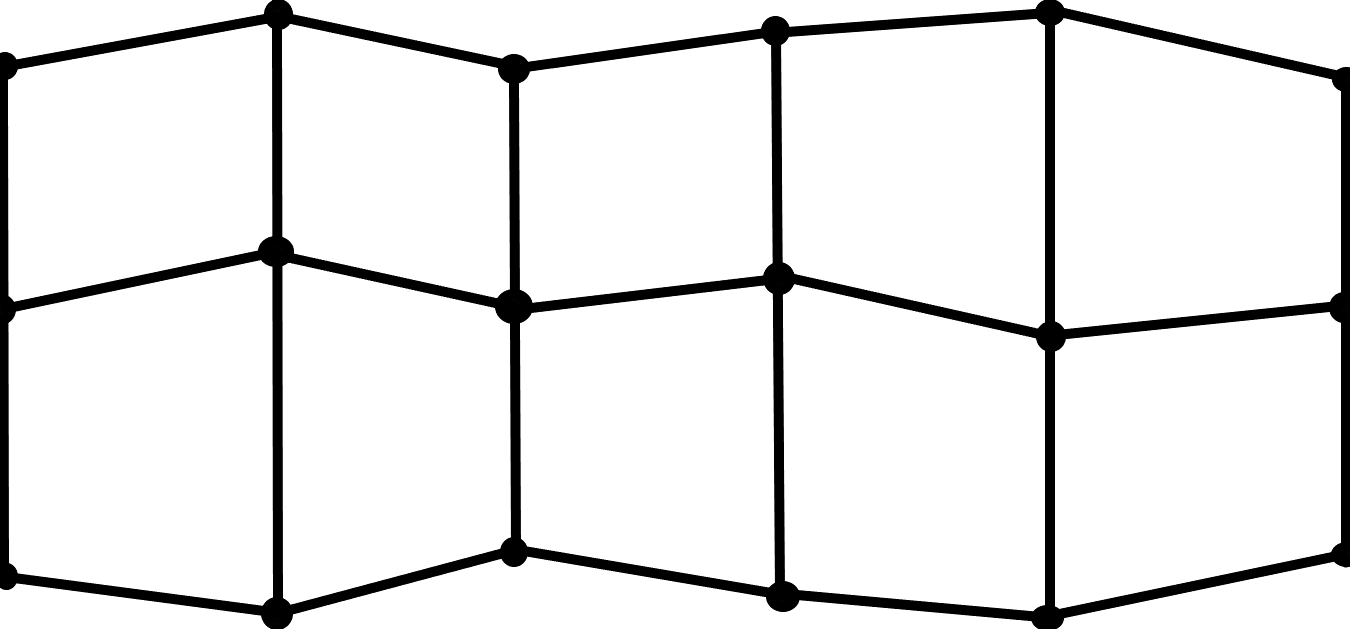}
  \caption{Post-shrinkage actual.}\label{fig:post-shrinkage actual}
\endminipage
\end{figure}

Additionally, we allow that the aspect ratio $\xi$ may vary as proteins denature. Based on a review of the literature, summarized in Table \ref{tab:shrink}, we conclude that if $\xi$ is initially $1$, a final value of $1.25$ is appropriate. We let $\xi$ vary sigmoidally with temperature (as $\chi_p$ does):
\begin{equation}
\frac{h_x}{h_y}=\xi:=1+\frac{0.25}{1+A\exp(\gamma(T-T_e))} \qquad h_x h_y=v
\label{eq:xi}
\end{equation}
This yields
\begin{equation}
h_x=h_0\sqrt{\frac{\phi_{init}\xi}{\phi}} \qquad h_y=h_0\sqrt{\frac{\phi_{init}}{\phi\xi}}.
\label{eq:h}
\end{equation}

\begin{table}[h]
\caption{Summary of shrinkage data from literature. Percent changes in each dimension are given. The L/T column gives final length/thickness ratio divided by the initial length/thickness ratio; W/T does the same for width/thickness.}
\begin{tabular}{l|c|c|c|c|c}
Cut & $\Delta$Length (\%) & $\Delta$Width (\%) & $\Delta$Thickness (\%) & L/T & W/T\\
\hline
Biceps femoris \cite{obuz} & $-0.35$ & $-12.21$ & $-25.61$ & 1.18 & 1.34\\
Biceps femoris \cite{ritchey} & $+6$ & $-1.8$ & $-19$ & 1.21 & 1.31\\
Longissimus lumborum \cite{obuz} & $-6.29$ & $-6.37$ & $-23.30$ & 1.22 & 1.22\\
Longissimus dorsi \cite{ritchey} & $-7.5$ & $+12$ & $-26.5$ & 1.43 & 1.18
\end{tabular}
\label{tab:shrink}
\end{table}

Because shrinkage is local and nonuniform, the cells that begin rectangular become irregular quadrangles and right angles become oblique (see Figures \ref{fig:pre-shrinkage}-\ref{fig:post-shrinkage actual}). However, in the limit of infinitesimal step size, local orthogonality is preserved. The notion is reminiscent of a Riemanian space with metric
\begin{equation}
 \mathbf{g} =
 	\begin{bmatrix} 
    \frac{\phi_{init}\xi}{\phi}&0\\
    0&\frac{\phi_{init}}{\phi\xi}
	\end{bmatrix}. 
\end{equation}
When calculating finite difference derivatives, we make the approximation of orthogonality, despite our finite step size.

We note that in reality, contraction of meat is mostly in the fiber direction only. It can happen that it expands in the orthogonal direction, as the diameter of the contracting fiber expands. We restrict this shrinkage model to isotropic deformation and leave anisotropic deformation for future work. 
\section{Numerical Method and Simulations}\label{Sec:Numerics}
\subsection{Discretized Equations}
We discretize the above model using finite differences. We approximate first-order time derivatives with a forward difference scheme, and both spatial first- and second-order derivatives with a central difference scheme. We define the discrete central difference operator $D_c$ on a function $f_{i}$ as the average of forward and backward differences, $D_f$ and $D_b$, respectively.
\[D_c\{f\} = \frac{1}{2}\Bigg[D_f\{f\} + D_b\{f\}\Bigg] = \frac{1}{2}\Bigg[\frac{f_{i+1}-f_{i}}{\frac{1}{2}[h_{i}+h_{i+1}]} + \frac{f_{i}-f_{i-1}}{\frac{1}{2}[h_{i}+h_{i-1}]}\Bigg]\]
We define the discrete second derivative operator $D^2$ as the difference between forward and backward difference first derivatives, divided by the step size.
\[D^2\{f\} = \frac{1}{h_i}\Bigg[\frac{f_{i+1}-f_i}{\frac{1}{2}[h_{i+1}+h_i]}-\frac{f_i-f_{i-1}}{\frac{1}{2}[h_i+h_{i-1}]}\Bigg]\]

\subsection{Numerical results}\hfill
\par
\begin{figure*}
        \begin{subfigure}[b]{0.475\textwidth}   
            \centering 
            \includegraphics[width=\textwidth]{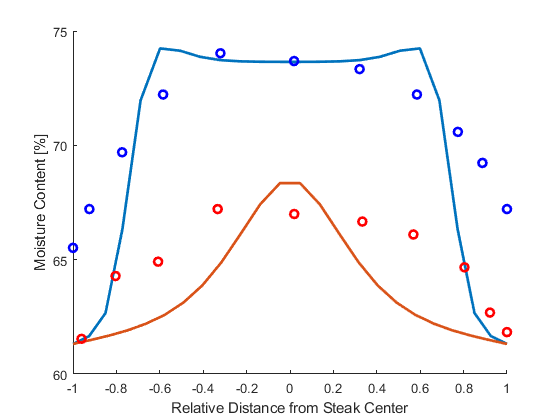}
            \caption[]%
            {{\small Moisture content for core temp $40^\circ C$ (blue), $70^\circ C$ (red).}}    
            \label{fig:bengston_moisture_comp}
        \end{subfigure}
        \quad
        \begin{subfigure}[b]{0.475\textwidth}   
            \centering 
            \includegraphics[width=\textwidth]{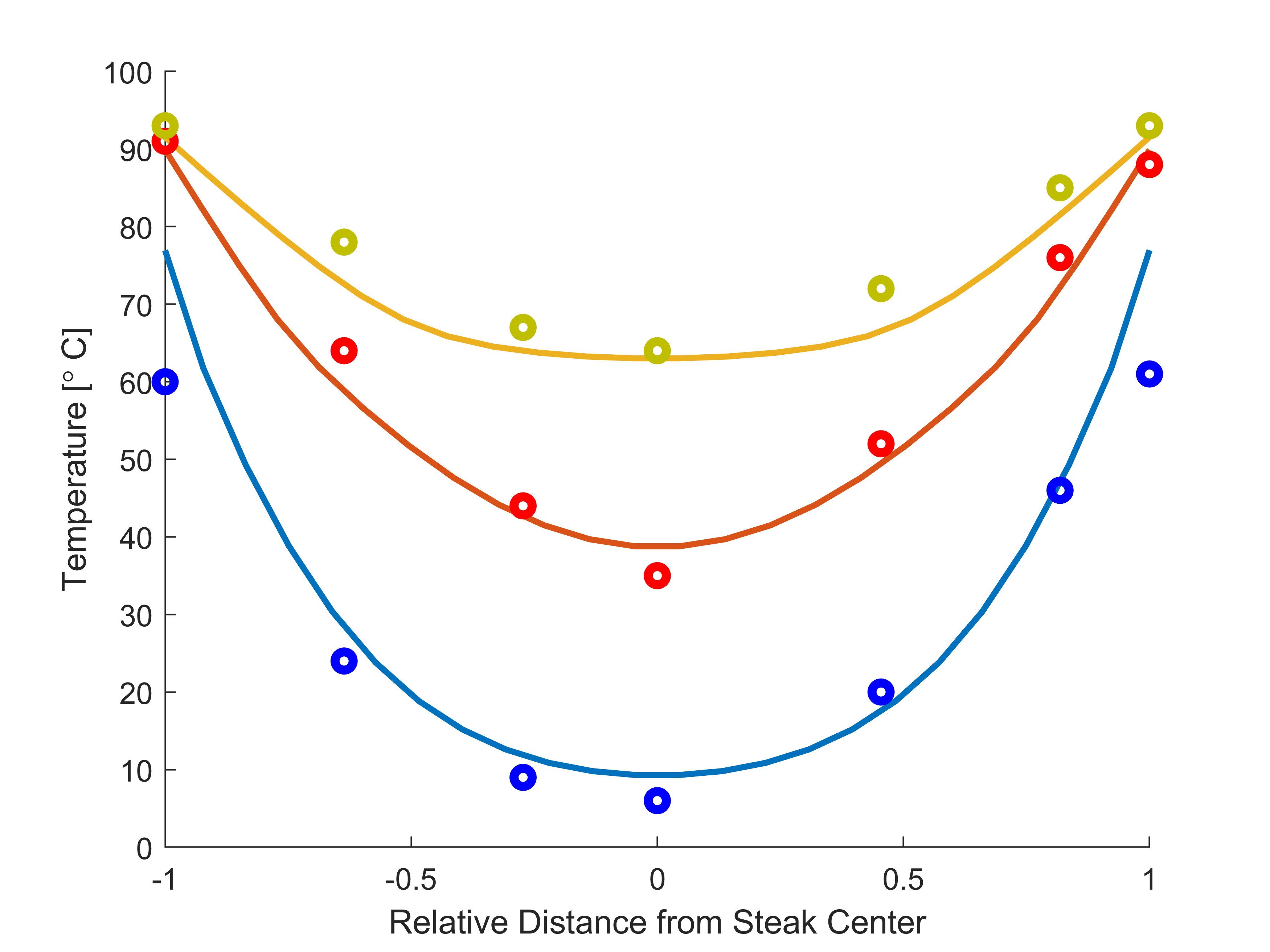}
            \caption[]%
            {{\small Temperature at 10 (blue), 30 (red), 50 (yellow) min.}}      
            \label{fig:bengston_temperature_comp}
        \end{subfigure}
                \caption{\small Comparison of simulations with empirical data from Bengston, et al \cite{Bengston}. $T_D = 225^\circ C$, $\kappa = 2.5\times 10^{-18}m^2\text{ to }2.5\times10^{-17} m^2$  (simulation (solid), empirical (o)).} 
         \label{Bengston_comparison}
\end{figure*}

\begin{figure*}
        \begin{subfigure}[b]{0.475\textwidth}   
            \centering 
            \includegraphics[width=\textwidth]{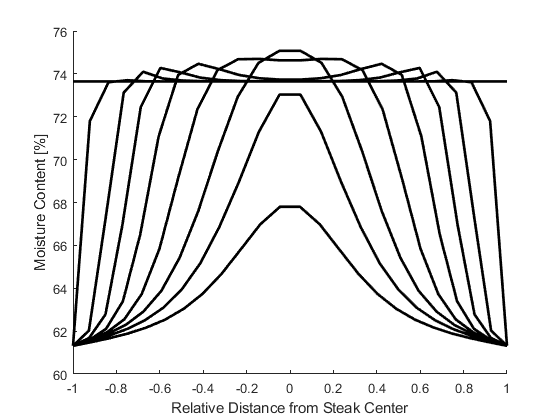}
            \caption[]%

            \label{fig:bengston_moisture_comp}
        \end{subfigure}
        \quad
        \begin{subfigure}[b]{0.475\textwidth}   
            \centering 
            \includegraphics[width=\textwidth]{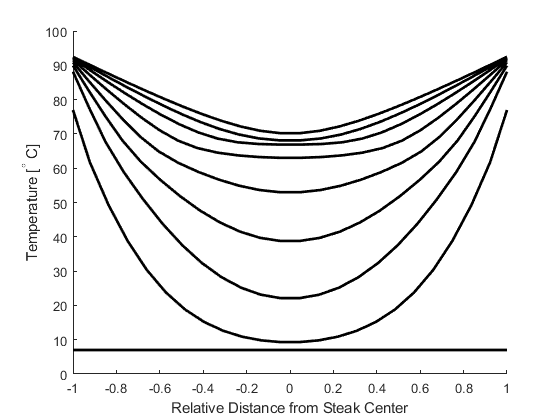}
            \caption[]%
      
            \label{fig:bengston_temperature_comp}
        \end{subfigure}
                \caption{\small Vertical Profiles of Moisture Content (A) and Temperature (B) at 10 minute intervals.} 
         \label{Bengston_comparison_10min}
\end{figure*}

 In the interest of validating the model with empirical data, we match the experimental conditions of Bengston, et al. \cite{Bengston} in oven roasting simulations --- namely, simulating oven roasting of an 5.5 x 8.0 cm steak at an oven temperature of $T_D=225^\circ C$ where the initial temperature of the steak is $T_0=7 ^\circ$C and environment vapor concentration $c_0$ is fit to experimental data (Figures \ref{Bengston_comparison}, \ref{Bengston_comparison_10min}). 

Simulations were performed using a grid of 24 x 35 points and a time-step of $2\times 10^{-4}$. This spatial resolution and time-step reflected reasonable convergence in comparison with more dense grids. 

Due to a lack of established permeability and water diffusion coefficient in literature, the simulation is produced using an initial permeability $\kappa$ of $2.5\times 10^{-18}m^2$ and a final permeability of $2.5\times 10^{-17}m^2$ sigmoidally fit in the same way as Florry-Huggins interaction parameter. 

We also considered simulations with constant permeability (even though in reality permeability is variable). Among all permeability values, a swelling of moisture is observed in the center of the steak, while the surface of the steak dries out. This swelling of moisture is most excessive in the low permeability simulations where moisture content rises by as much as $10\%$ which is inconsistent with empirical measurements of swelling on the order of $1\%$; thus, we conclude that $\kappa = 10^{-16}m^2$ is a more accurate description of moisture content dynamics. Moreover, in low permeability simulations, moisture content sometimes reaches $100\%$, meaning $\phi=0$. We can interpret the excessive moisture content increase as a result of the changing slope of $\pi_{sw}$ (Figure \ref{fig:pi_sw}). If the temperature of the meat increases faster than moisture content can decrease, then the slope of $\pi_{sw}$ with respect to $\phi$ changes from monotonically negative to positive allowing $\phi$ to drift towards the other root of $\pi_{sw}$, which is at $\phi = 0$. By simulating the full non-linear behavior of $\pi_{sw}$ rather than a linearization, it is clear that a low permeability of $\kappa = 10^{-17}m^2$ is not physically correct since it predicts unrealistically high moisture content increases. 

Our model is validated with empirical data showing rough agreement between simulation and experiment (Figures \ref{Bengston_comparison}, \ref{Bengston_comparison_10min}). 
\begin{figure*}
                \begin{subfigure}[b]{0.425\textwidth}   
            \centering 
            \includegraphics[width=\textwidth]{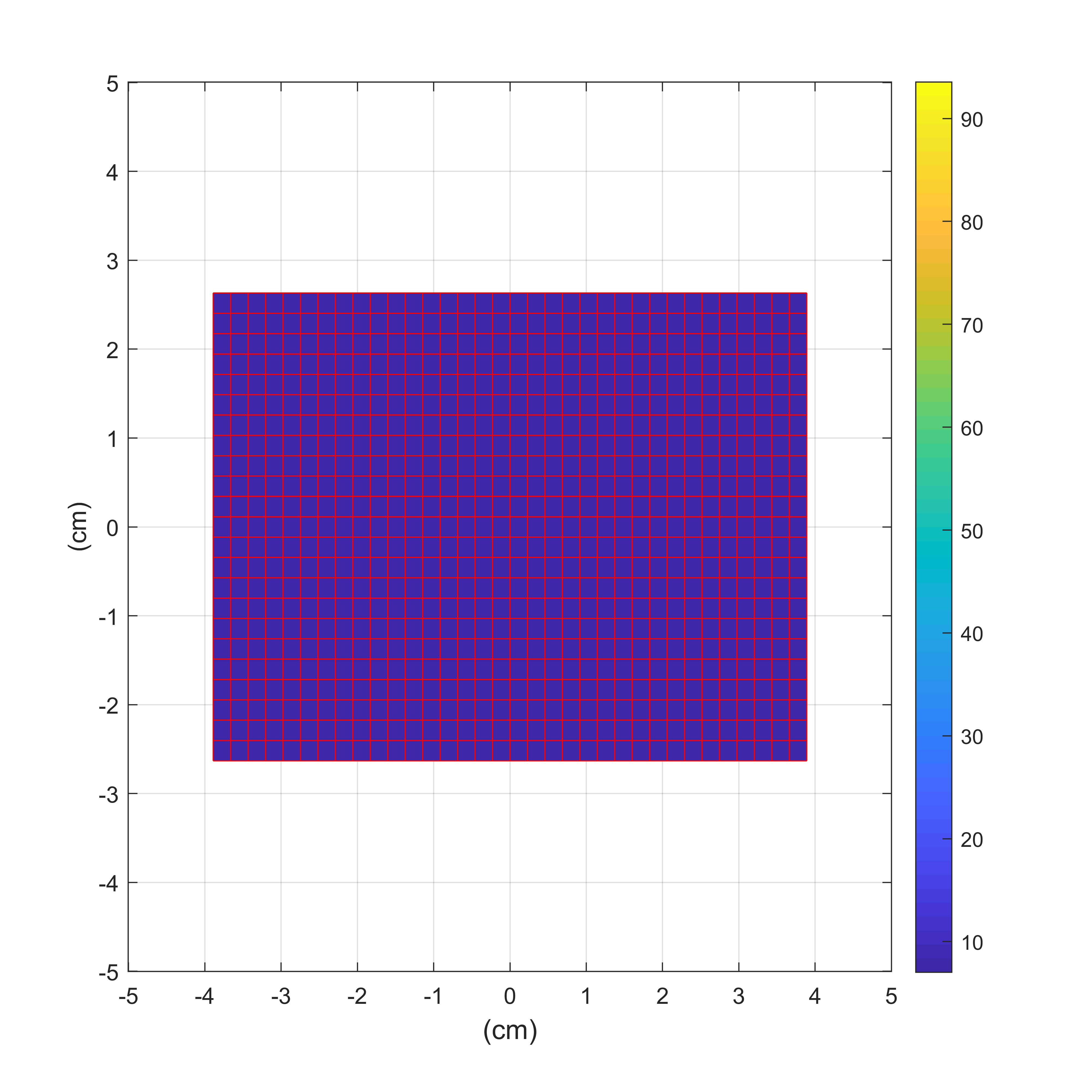}
            \caption[]%
            {{\small $0$ min}}    
        \end{subfigure}
        \quad
        \begin{subfigure}[b]{0.425\textwidth}   
            \centering 
            \includegraphics[width=\textwidth]{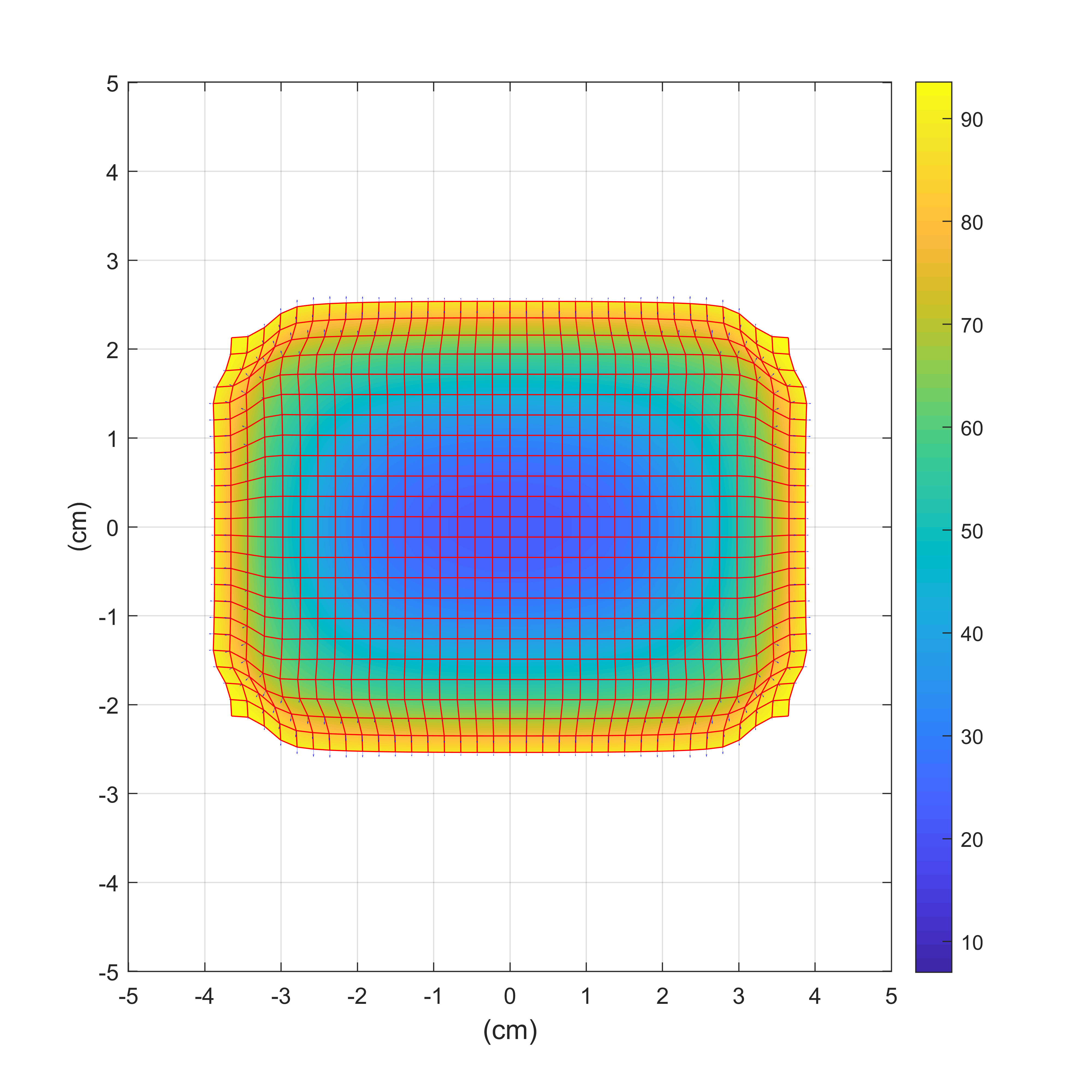}
            \caption[]%
            {{\small $20$ min}}    
        \end{subfigure}
        \begin{subfigure}[b]{0.425\textwidth}   
            \centering 
            \includegraphics[width=\textwidth]{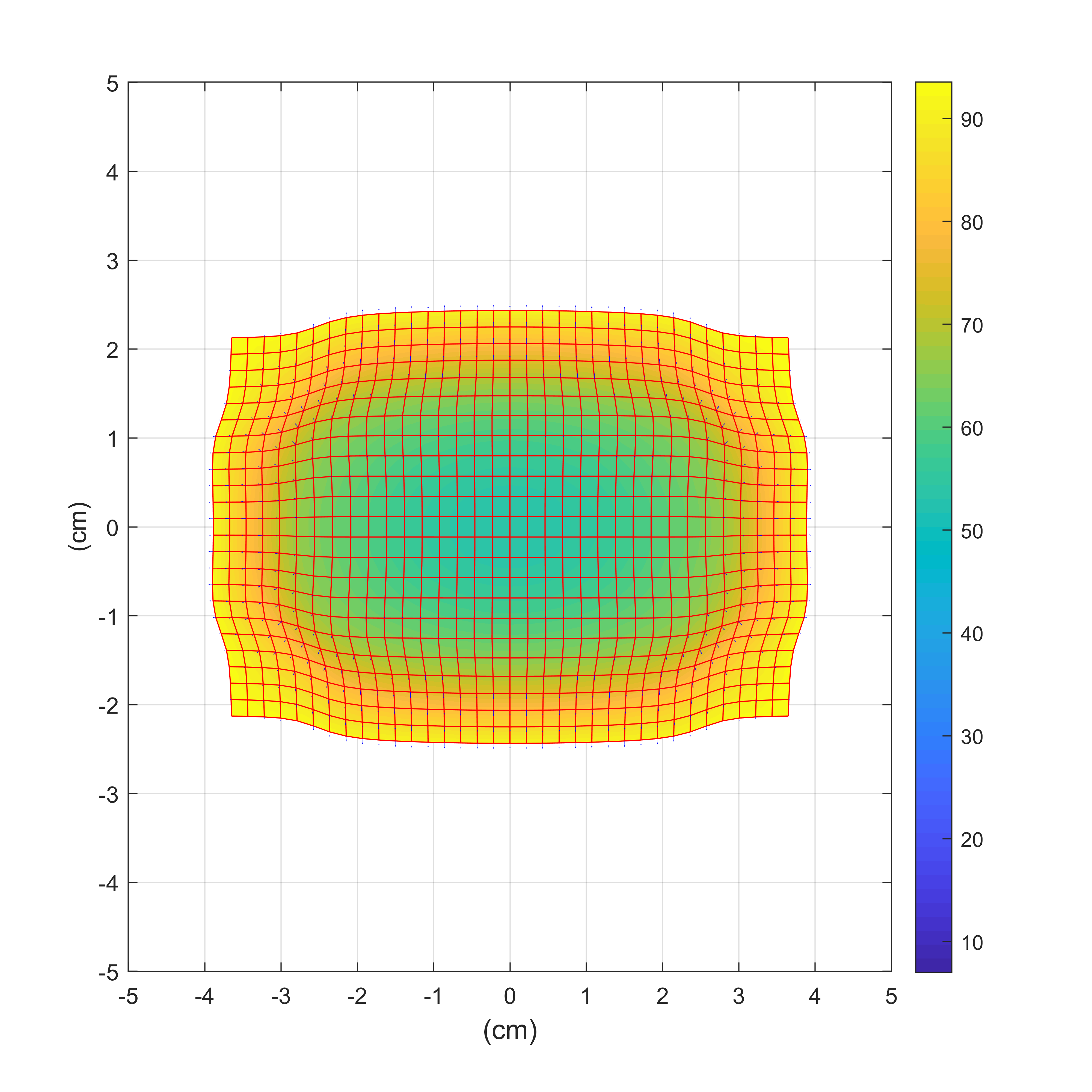}
            \caption[]%
            {{\small $40$ min}}    
        \end{subfigure}
        \quad
        \begin{subfigure}[b]{0.425\textwidth}   
            \centering 
            \includegraphics[width=\textwidth]{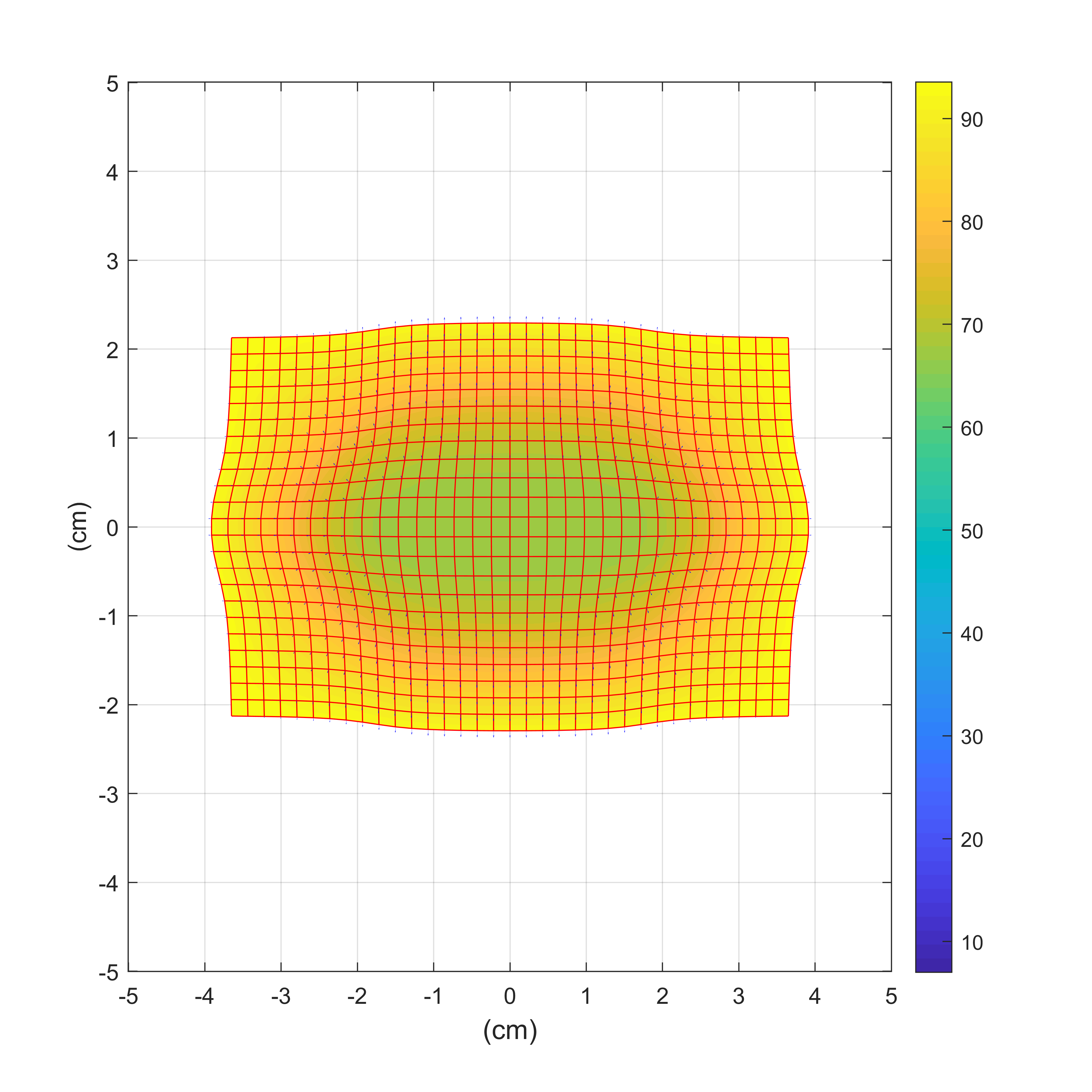}
            \caption[]%
            {{\small $60$ min}}    
        \end{subfigure}
        \caption[The average and standard deviation of critical parameters ]
        {\small Heat Map and deformed shape at 20 min intervals for oven roasting $T_D = 225^\circ C$} 
        \label{shrinkage_shape}
\end{figure*}

\begin{figure*}
        \begin{subfigure}[b]{0.475\textwidth}   
            \centering 
            \includegraphics[width=\textwidth]{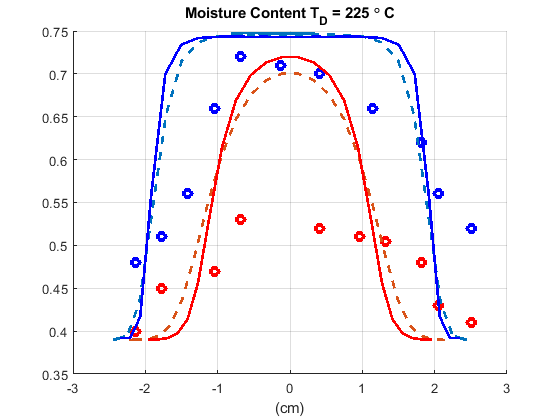}
            \caption[]%

            \label{fig:bengston_moisture_comp16}
        \end{subfigure}
        \quad
        \begin{subfigure}[b]{0.475\textwidth}   
            \centering 
            \includegraphics[width=\textwidth]{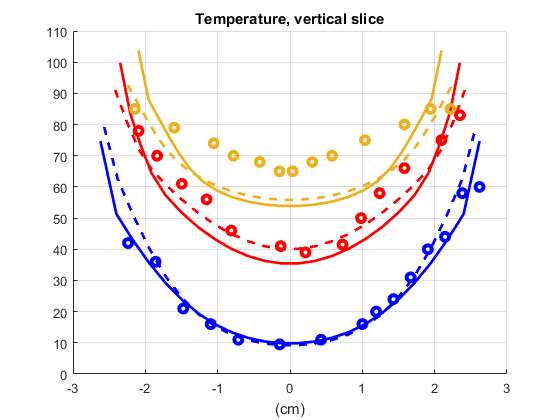}
            \caption[]%
      
            \label{fig:bengston_temperature_comp16}
        \end{subfigure}
                \caption{\small Comparison of simulations with empirical data from Bengston, et al \cite{Bengston}. $T_D = 225^\circ C$, $\kappa = 10^{-16} m^2$ (simulation (solid), empirical (o)). Even though this was the best among constant permeability values, it obviously does not perform as well as the case for variable permeability.}
         \label{Bengston_comparison_16}
\end{figure*}

        \pagebreak


\section{Conclusions}
Our model differs from its predecessors in its preservation of fully nonlinear equations, as opposed to the linearization seen in \cite{van2007soft} and others, and in incorporating two-dimensional local shrinkage. We demonstrate reasonable agreement with empirical data. The model captures the swelling of moisture in the steak center and the drying out at the surface that is observed during a steak cooking process. 

When using variable permeability, we find that $\kappa = 2.5\times 10^{-18}\text{ to }2.5\times10^{-17} m^2$ produces the best agreement with empirical data, which agrees with the value suggested in \cite{Feyissa}. In the simplified case where the permeability is constant, we found compelling evidence for a higher value of $\kappa \sim 10^{-16}$ $m^2$. Hence using variable permeability gives a better and more realistic model.

As for the water diffusion constant $D_w$, we find that it has a smoothing effect on poorly behaved simulations, but by no means did it fully rectify the behavior of these simulations. Its impact is modest on well-behaved simulations and we are in favor of keeping $D_w=0$ as presented in the paper.

The model extends easily to three dimensions, however it will be more computationally expensive and it will be the subject of future work. 

\section{Acknowledgements}
This work was completed in the summer of 2018 as part of the Research Experiences for Undergraduate (REU) program at James Madison University, funded by National Science Foundation grant NSF-DMS 1560151.

\appendix
\section{Discretization}
\subsection{Equations}
We define the derivative operators $D_x$, $D_y$, $D_{xx}$, and $D_{yy}$ on a function of two variables $f_{i,j}$ as
\begin{align*}
D_{x}f_{i,j} &= \frac{f_{i+1,j}-f_{i,j}}{h^x_{i+1,j}+h^x_{i,j}} + \frac{f_{i,j}-f_{i-1,j}}{h^x_{i,j}+h^x_{i-1,j}}
& &
D_{y}f_{i,j} = \frac{f_{i,j+1}-f_{i,j}}{h^y_{i,j+1}+h^y_{i,j}} + \frac{f_{i,j}-f_{i,j-1}}{h^y_{i,j}+h^y_{i,j-1}}
\\
D_{xx}f_{i,j} &= \frac{2}{h^x_{i,j}}\bigg[\frac{f_{i+1,j}-f_{i,j}}{h^x_{i+1,j}+h^x_{i,j}} - \frac{f_{i,j}-f_{i-1,j}}{h^x_{i,j}+h^x_{i-1,j}}\bigg]
& &
D_{yy}f_{i,j} = \frac{2}{h^y_{i,j}}\bigg[\frac{f_{i,j+1}-f_{i,j}}{h^y_{i,j+1}+h^y_{i,j}} - \frac{f_{i,j}-f_{i,j-1}}{h^y_{i,j}+h^y_{i,j-1}}\bigg].
\end{align*}
We obtain the following systems of equations:
\begin{enumerate}
\item \textbf{Porosity-fluid velocity system of equations.} 
Discretizing the continuity equation \eqref{eqn:continuity} yields
\begin{equation*}
\phi^{k+1}_{i,j} = \phi^k_{i,j} + \Delta t\left[D_x \{(1-\phi_{i,j})w^{(1)}_{i,j}\}+ D_y \{(1-\phi_{i,j})w^{(2)}_{i,j}\}+\frac{D_w t_0}{l^2} \left(D_{xx} \{\phi\}+D_{yy}\{\phi\} \right) \right].
\end{equation*}
\item \textbf{Temperature-porosity-fluid velocity system of equations.} Next, we discretize the dimensionless swelling, elastic, and mixing pressures in \eqref{eqn:pi_mix} and \eqref{eqn:pi_el}:
\begin{align*}
\pi_{sw_{i,j}}^k&=\pi_{el_{i,j}}^k+\pi_{mix_{i,j}}^k\\
\pi_{el_{i,j}}^k&=\beta_{el} \bigg(1+\frac{\nu}{\alpha} T_{i,j}^k\bigg) \bigg[ {\phi_{i,j}^{k}}^{1/3}\phi_0^{2/3}-\frac{1}{2}\phi_{i,j}^k\bigg], \\
\pi_{mix_{i,j}}^k&=\beta_{mix} \bigg(1+\frac{\nu}{\alpha} T_{i,j}^k\bigg) \bigg[\ln(1-\phi_{i,j}^k)+\phi_{i,j}^k + \chi_{i,j}^k\left({\phi_{i,j}^{k}}\right)^2\bigg].
\end{align*}
Similarly, we discretize the Flory-Huggins parameters in \eqref{eqn:FH_interation} and \eqref{eqn:FH_sigmoidal},
\begin{align*}
\chi(T,\phi)_{i,j}^k &= \chi_{p_{i,j}}^k - \left(\chi_{p_{i,j}}^k-\chi_0\right)\left(1-\phi_{i,j}^k\right)^2,
\\
\chi_p(T)_{i,j}^k &= \chi_{pn}-\frac{\chi_{pd}-\chi_{pn}}{1+A\exp\left[\gamma\left(T_{i,j}^k\left(T_D-T_0\right)+T_0-T_e\right)\right]}. 
\end{align*}
We discretize the dimensionless dynamic viscosity of the fluid $\mu(T)$:
\begin{align*}
\mu_{i,j}^k = \frac{2.42\times10^{-5}}{\mu(T_0)}\left(10^{\frac{247.8}{T_{i,j}^k(T_D-T_0)+T_0-140}}\right).
\end{align*}
We discretize the dimensionless momentum equation \eqref{eqn:momentum}. The $x$-component gives
\begin{align*}
(1-\phi^k_{i,j})w^{(1),k}_{i,j}
= \frac{-\kappa^{(11)}}{\mu^k_{i,j}}D_x\{\pi^k_{sw_{i,j}}\},
\end{align*}
while the $y$-component yields
\begin{align*}
(1-\phi^k_{i,j})w^{(2),k}_{i,j}
= \frac{-\kappa^{(22)}}{\mu^k_{i,j}}D_y\{\pi^k_{sw_{i,j}}\}.
\end{align*}
\item \textbf{Fluid velocity-temperature-porosity system of equations.}
Discretization the left side of the non-dimensional energy balance equation \eqref{eqn:AdvDiff} gives
\begin{align*}
\left(1-\phi^k_{i,j}(1-\nu)\right)\left(\frac{T^{k+1}_{i,j}-T^k_{i,j}}{\Delta t}\right) + \left(\alpha+\nu T^k_{i,j}\right)\left(\frac{\phi^{k+1}_{i,j}-\phi^k_{i,j}}{\Delta t}\right) + \left(1-\phi^k_{i,j}\right)\left[w^{(1),k}_{i,j}D_x\{T^k_{i,j}\}+w^{(2),k}_{i,j}D_y\{T^k_{i,j}\}\right]\\
-\frac{D_{w}}{D}\bigg(T_{i,j}^k+\frac{\alpha}{\nu}\bigg)\bigg(D_{xx}{\phi_{i,j}^k}+D_{yy}{\phi_{i,j}^k}\bigg).
\end{align*}
The discretization of the right side of \eqref{eqn:AdvDiff} is
\begin{align*}
\nabla \cdot \left(k \nabla T\right) & = \left(\frac{\omega^2}{\left(\omega(1-\phi)+1\right)^2}D_x\{\phi_{i,j}\}D_x\{T^k_{i,j}\}+\frac{\omega}{\omega(1-\phi)+1}D_{xx}\{T^k_{i,j}\}\right) + \\ &
\qquad \Bigg(\left(\omega-1\right)D_y\{\phi_{i,j}\}D_y\{T^k_{i,j}\}+\left(1-\phi\left(1-\omega\right)\right)D_{yy}\{T^k_{i,j}\}\Bigg). \\
\end{align*}
\end{enumerate}
\vspace{-2em}\subsection{Discretized boundary conditions} 
Porosity and energy balance boundary conditions are a coupled non-linear system. To avoid the computational expense of solving the non-linear system each time-step; instead, we solve the full non-linear system for the first time step only, and use the previous time step values to approximate non-linear and coupled terms. We discretize on a grid of $\theta N$ by $N$ points, initially with uniform spacing.
\begin{enumerate}
	\item \textbf{The Dirichlet boundary condition for the porosity}:
    \[\pi_{sw_{1,j}}^k(T_{1,j}^{k-1},\phi_{1,j}^k)=\pi_{sw_{\theta N,j}}^k(T_{\theta N,j}^{k-1},\phi_{\theta N,j}^k)=\pi_{sw_{i,1}}^k(T_{i,1}^{k-1},\phi_{i,1}^k)=\pi_{sw_{i,N}}^k(T_{i,N}^{k-1},\phi_{i,N}^k)=0.\]
    
    \item \textbf{Momentum boundary conditions}.
    Taking $\pi_{sw}=0$ on the boundary, and using the appropriate one-sided derivative on the boundary, we have the following:
    \begin{enumerate}
    	\item On the left: 
        \begin{align*}
(1-\phi^k_{1,j})w^{(1),k}_{1,j}
= \frac{\kappa^{(11)}}{\mu^k_{1,j}}\left(\frac{\pi_{sw_{2,j}}^k}{h^x_{2,j}+h^x_{1,j}}\right), \qquad \qquad w^{(2),k}_{1,j}=0
\end{align*}
        \item On the right:
        \begin{align*}
(1-\phi^k_{\theta N,j})w^{(1),k}_{\theta N,j}
= \frac{-\kappa^{(11)}}{\mu^k_{\theta N,j}}\left(\frac{\pi_{sw_{\theta N-1,j}}^k}{h^x_{\theta N-1,j}+h^x_{\theta N,j}}\right), \quad w^{(2),k}_{\theta N,j}=0
\end{align*}
        
        \item On the top:
\begin{align*}
(1-\phi^k_{i,1})w^{(2),k}_{i,1}
= \frac{\kappa^{(22)}}{\mu^k_{i,1}}\left(\frac{\pi_{sw_{i,2}}^k}{h^y_{i,1}+h^y_{i,2}}\right), \qquad \quad \qquad w^{(1),k}_{i,1}=0.
\end{align*}
        
        \item On the bottom:
\begin{align*}
(1-\phi^k_{i,N})w^{(2),k}_{i,N}
= \frac{-\kappa^{(22)}}{\mu^k_{i,N}}\left(\frac{\pi_{sw_{i,N-1}}^k}{h^y_{i,N}+h^y_{i,N-1}}\right), \qquad \quad w^{(1),k}_{i,N}=0.
\end{align*}
    \end{enumerate}
    
    \item \textbf{Temperature boundary conditions}. $T^{k-1}_{i,j}$ is used as an approximation of $T^k_{i,j}$ in the non-linear $j_{evap}$ term and $T^{k-1}_{i,j}$ is used to approximate to avoid solving the coupled non-linear system of the energy boundary condition and the porosity boundary condition simultaneously. This approximation is valid for small $\Delta t$.

    \begin{enumerate}
    	\item At the left boundary,
        \begin{align*}-\left(\frac{1}{\omega(1-\phi_{1,j}^{k-1})+\phi_{1,j}^{k-1}}\right)\frac{T_{1,j}^k-T_{2,j}^k}{\frac{1}{2}(h^{x,k}_{1,j}+h^{x,k}_{2,j})} & =       T_{1,j}^k-1+\lambda j_{evap}\left(\phi_{1,j}^{k-1},T_{1,j}^{k-1}\right).
        \end{align*}
            	\item At the right boundary,
        \begin{align*}-\left(\frac{1}{\omega(1-\phi_{\theta N,j}^{k-1})+\phi_{\theta N,j}^{k-1}}\right)\frac{T_{\theta N,j}^k-T_{\theta N -1,j}^k}{\frac{1}{2}(h^{x,k}_{\theta N,j}+h^{x,k}_{\theta N-1,j})} & =       T_{1,j}^k-1+\lambda j_{evap}\left(\phi_{\theta N,j}^{k-1},T_{\theta N,j}^{k-1}\right).
        \end{align*}
        
              \item At the top boundary,
        \begin{align*}-\left(\frac{1-\phi_{i,1}^{k-1}}{\omega}+\phi_{i,1}^{k-1}\right)\frac{T_{i,1}^k-T_{i,2}^k}{\frac{1}{2}(h^{y,k}_{i,1}+h^{y,k}_{i,2})} & =       T_{i,1}^k-1+\lambda j_{evap}\left(\phi_{i,1}^{k-1},T_{i,1}^{k-1}\right).
        \end{align*}
        
            \item At the bottom boundary,
        \begin{align*}-\left(\frac{1-\phi_{i,N}^{k-1}}{\omega}+\phi_{i,N}^{k-1}\right)\frac{T_{i,N}^k-T_{i,N-1}^k}{\frac{1}{2}(h^{y,k}_{i,N}+h^{y,k}_{i,N-1})} & =       T_{i,N}^k-1+\lambda j_{evap}\left(\phi_{i,N}^{k-1},T_{i,N}^{k-1}\right).
        \end{align*}
    \end{enumerate}
\end{enumerate}

\subsection{Pseudo-code} Our code proceeds in the following order:
  \begin{lstlisting}[mathescape=true]
Require: $\phi^k$, $T^k$, $w^k$, $h^k$
	1. Continuity:							$\phi^{k+1}$ on bulk $\Omega$ $\leftarrow$  $\phi^k$, $\mathbf{w}^{k}$, $h^k$.
	2. Energy Balance:					$T^{k+1}$ on bulk $\Omega$ $\leftarrow$ $\phi^k$, $T^{k}$, $h^k$, and $\phi^{k+1}$ on bulk $\Omega$.
	3. Shrinkage: 							$h^{k+1}$  on bulk $\Omega$ $\leftarrow$ $\phi^{k+1}$ on bulk $\Omega$
	4. Coupled Nonlinear B.C.s:		$T^{k+1}$, $\phi^{k+1}$  on $\partial\Omega$ $\leftarrow$ $T^{k+1}$, $\phi^{k+1}$, $h^{k+1}$ on bulk $\Omega$.
	5. Shrinkage: 							$h^{k+1}$  on $\partial \Omega$ $\leftarrow$ $\phi^{k+1}$ on $\partial \Omega$
	6. Momentum Balance: 				$\mathbf{w}^{k+1}$ on $\partial \Omega$ and bulk $\Omega$ $\leftarrow$  $\phi^{k+1}$, $T^{k+1}$, $h^{k+1}$ on $\partial \Omega$ and bulk $\Omega$.    
  \end{lstlisting}


\end{document}